\documentclass[11pt,letterpaper]{article}
\usepackage{graphicx}
\usepackage{tikz}
\usepackage{verbatim}
\usepackage{wrapfig}
\usepackage{url}
\usepackage{cite}
\usepackage{xcolor}
\usepackage{subcaption}
\usepackage{placeins}
\begin{document}

\title{\vspace{-.725in}\Large\bf 
Letter of Intent: Muonium R\&D/Physics Program at the MTA\vspace{-.1in}}

\author{\hspace{-.25in} C. Gatto,$^{5,6}$ C. Izzo,$^2$ C. J. Johnstone,$^2$ D. M. Kaplan,\footnotemark[1]~\,$^{3,4}$ K. R. Lynch,$^2$ \cr 
\hspace{-.25in}D. C. Mancini,$^3$  A. Mazzacane,$^2$ B. McMorran,$^7$ J. P. Miller,$^1$ 
J. D. Phillips,$^3$\footnotemark[2]  \cr 
\hspace{-.25in} T. J. Phillips,$^3$ R. D. Reasenberg,$^8$ T. J. Roberts,$^{3,4}$ J. Terry$^3$
\\[0.125in]
{\it \hspace{-.25in}$^1$Boston U., $^2$Fermilab, $^3$Illinois Institute of Technology, $^4$Muons, Inc., $^5$INFN Napoli, }\cr 
{\it \hspace{-.25in}$^6$Northern Illinois U., $^7$U. Oregon, $^8$U. California San Diego CASS
}
\\[0.1in]
\leftline{\small\footnotemark[1]~~Spokesperson~~~~~  \footnotemark[2]~Also at Zurich Instruments}
\date{December 5, 2022}
}

\maketitle\vspace{-.5in}

\vspace{-.15in}
\section{Introduction}\vspace{-.05in}

There is a need for a high-efficiency source of muonium (M~$\equiv\mu^+ e^-$, chemically a light isotope of hydrogen), traveling as a beam in vacuum, for fundamental muon measurements, sensitive searches for symmetry violation, and precision tests of theory~\cite{Petrov}. Currently PSI in Switzerland is the world leader for such research. With PIP-II Fermilab has the potential to eclipse PSI and become the new world leader. It is prudent to begin the R\&D now in order to be ready when PIP-II comes online. Fermilab's MeV Test Area (MTA) at the 400 MeV H$^-$ Linac has a low-energy muon beamline suitable for this R\&D, with the potential to compete with PSI for this physics in the pre-PIP-II near term as well.

Key muonium measurements include the search for M-$\overline{\rm M}$ conversion, precision measurement of the M atomic spectrum, and the study of antimatter gravity using M. Furthermore, the J-PARC $g-2$ experiment proposes to use a low-energy $\mu^+$ beam produced by photo-ionizing a slow beam of muonium, but the needed high-intensity muonium beam has yet to be demonstrated. The technique we propose may form a suitable muonium source for such a $g-2$ measurement as well as for other applications of slow muon beams.

M-$\overline{\rm M}$ conversion is a double charged-lepton flavor-violating (CLFV) reaction, allowed (albeit at an undetectably small rate) via neutrino mixing. It may be no less likely\,---\,and in some models, {\it more} likely\,---\,than $\mu$ to $e$ conversion~\cite{Petrov}. Thus in a thorough CLFV research program it should be studied as well as Mu2e. The best current limit, $P_{
\rm M\overline{M}}\le 8.3\times10^{-11}$
(90\% C.L.) in 0.1\,T field~\cite{Willmann}, was published over 20 years ago and, given the technical progress since then, is ripe for reexamination.

As a pure QED bound state of two point-like particles, muonium offers a more direct test of theory than hydrogen, free of hadronic and finite-size effects. Precise predictions and measurements have been made of its 1S--2S~\cite{Meyer} and hyperfine~\cite{Liu} splittings  (to 4 and 12\,ppb, respectively), again over 20 years ago, and experiments are now under way at PSI and J-PARC to improve them.

Antimuon gravity has never been measured but its measurement now appears feasible, thanks to the new approach~\cite{Antognini} described below. While the weak equivalence principle (WEP) of general relativity implies that all forms of matter should act identically in a gravitational field, and precision measurements supporting it have been made using torsion pendula, the Earth--Moon--Sun system, and levitated cylinders in earth orbit~\cite{EotWash-LLR-MICROSCOPE}, it has been argued that the WEP may not hold for antimatter~\cite{NG}. Moreover, in theories that assume maximal WEP violation by antimatter (in which the gravitational acceleration of antimatter on earth, $\overline{g}$, satisfies $\overline{g}=-g$), major puzzles of cosmology can be resolved with no further assumptions and no need of the as-yet-unobserved dark matter and dark energy~\cite{Chardin}. Or a 5th force coupling non-universally to leptons, as suggested by muon $g - 2$ and $B$-decay anomalies~\cite{Aaij}, might cause $\overline{g}$ and $g$ to differ slightly.

\vspace{-.15in}\section{Approach}\vspace{-.05in}

Our method (proposed by PSI's D. Taqqu) relies on the efficient conversion of positive muons to muonium atoms in superfluid helium~\cite{Abela} combined with the predicted expulsion of muonium atoms at the superfluid surface due to the predicted large positive chemical potential of muonium in superfluid helium~\cite{Taqqu,Luppov}. An electric field maintained in the superfluid helium by means of an electron pool at the surface will drift stopped $\mu^+$ to the surface, where they will combine with electrons to form muonium and be expelled into the vacuum. 

\vspace{-.15in}\section{Apparatus, R\&D Tasks, Budget}\vspace{-.05in}

\begin{figure}[t]
\begin{center}
\begin{tikzpicture}
\begin{scope}
    \node {\includegraphics[height=2.25in, trim=15 0 0 0]{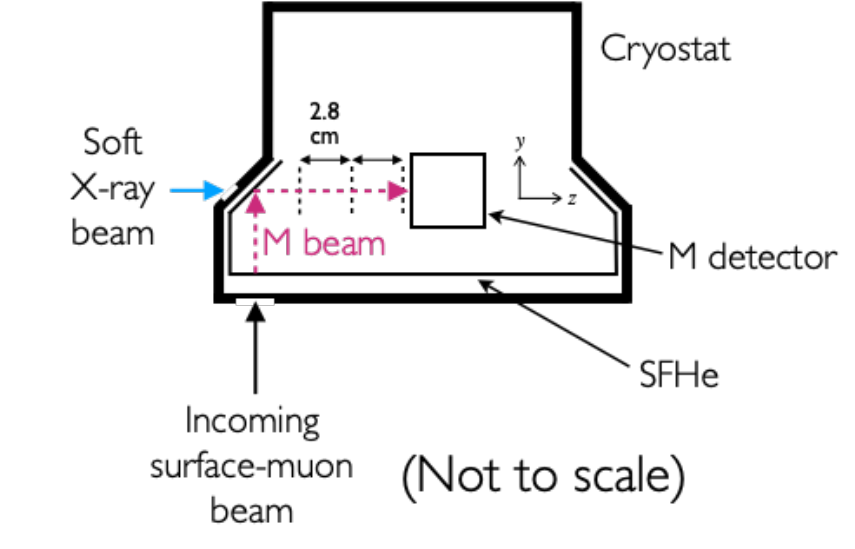}};
    \end{scope}
    \begin{scope}[xshift=3.1in, yshift=0.1in]
    \node {\includegraphics[height=2in,trim=0 0 0 0,clip]{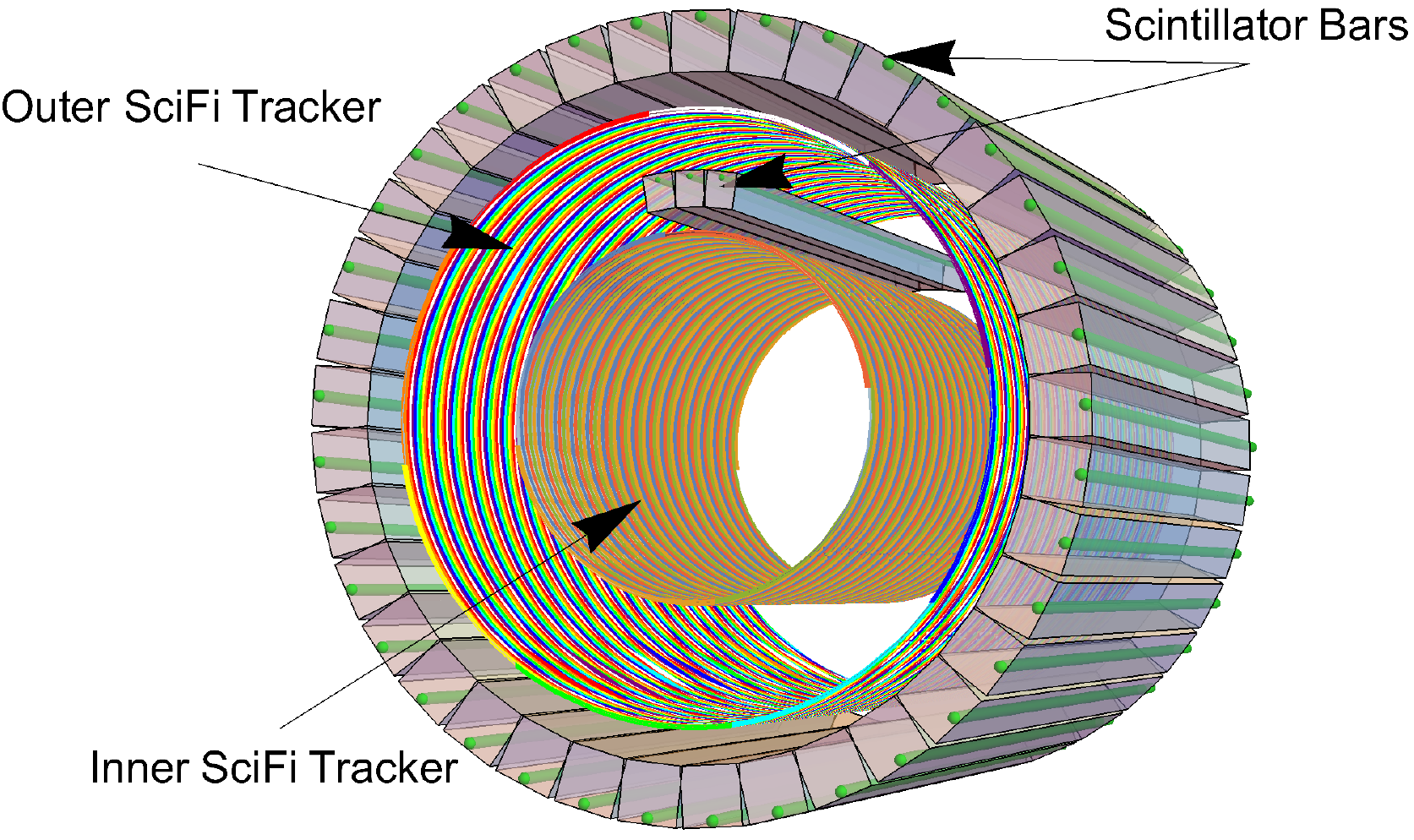}}; 
    \end{scope}
\end{tikzpicture}\vspace{-.15in}
\caption{\small (left) Conceptual sketch of ``Phase III" muonium-gravity apparatus: degraded surface muons stop in SFHe layer, producing upward-directed M beam, deflected into horizontal by SFHe-coated reflector~\cite{Luppov} and entering 3-grating interferometer; calibration soft X-ray beam enters from left; (right) muonium detector concept: fast decay $e^+$ detected in barrel SciFi detector triggered by scintillating-bar barrel hodoscope, slow $e^-$ in $xy$ scintillator hodoscope array with SiPM readout after electrostatic acceleration ($xy$ hodoscope and accelerating rings not shown).}
\label{fig:MAGE-app}
\end{center}\vspace{-.2in}
\end{figure}

Of the three key muonium measurements described above, muonium gravity is the one that has never been measured, uses the smallest apparatus, and will clearly fit within the MTA; the others require more evaluation and may have to await PIP-II. The key pieces of apparatus for our proposed program are then (1) a small cryostat that can be operated at $\sim 0.1$\,K, within which a pool of superfluid helium  (SFHe) can be created and maintained; the depth of the pool will be $\sim$\,100\,$\mu$m, sufficient to stop an appreciable fraction of the slow $\mu^+$ beam incident from below after its energy has been degraded by traversal of the cryostat wall; (2) a 3-grating atom interferometer with $\approx$\,100\,nm grating pitch and few-cm grating separation along $z$; and (3) an efficient detector of both the M decay products and a calibration X-ray beam; beams enter through thin spots or beryllium windows in the cryostat walls. The apparatus is shown in Fig.~\ref{fig:MAGE-app} and will be installed within the sample volume of a dilution refrigerator, operated at $\sim$\,0.1\,K to reduce the SFHe vapor pressure and suppress M-He scattering. Note that the actual grating separation will be determined after a thorough optimization study; Fig.~\ref{fig:MAGE-app}(left) shows a separation corresponding to 2 muon lifetimes at the predicted M velocity of 6.3\,mm/$\mu$s. Simulations show that the sign of $\overline{g}$ can be determined to 5$\sigma$ in a few days of running at $10^5$\,Mu/s  incident on the interferometer; if systematics can be adequately controlled, a 10\% measurement can be made in about a month.

\begin{wrapfigure}{R}{0.45\textwidth}
\vspace{-.9in}
\begin{center}
    \includegraphics[height=1.5in]{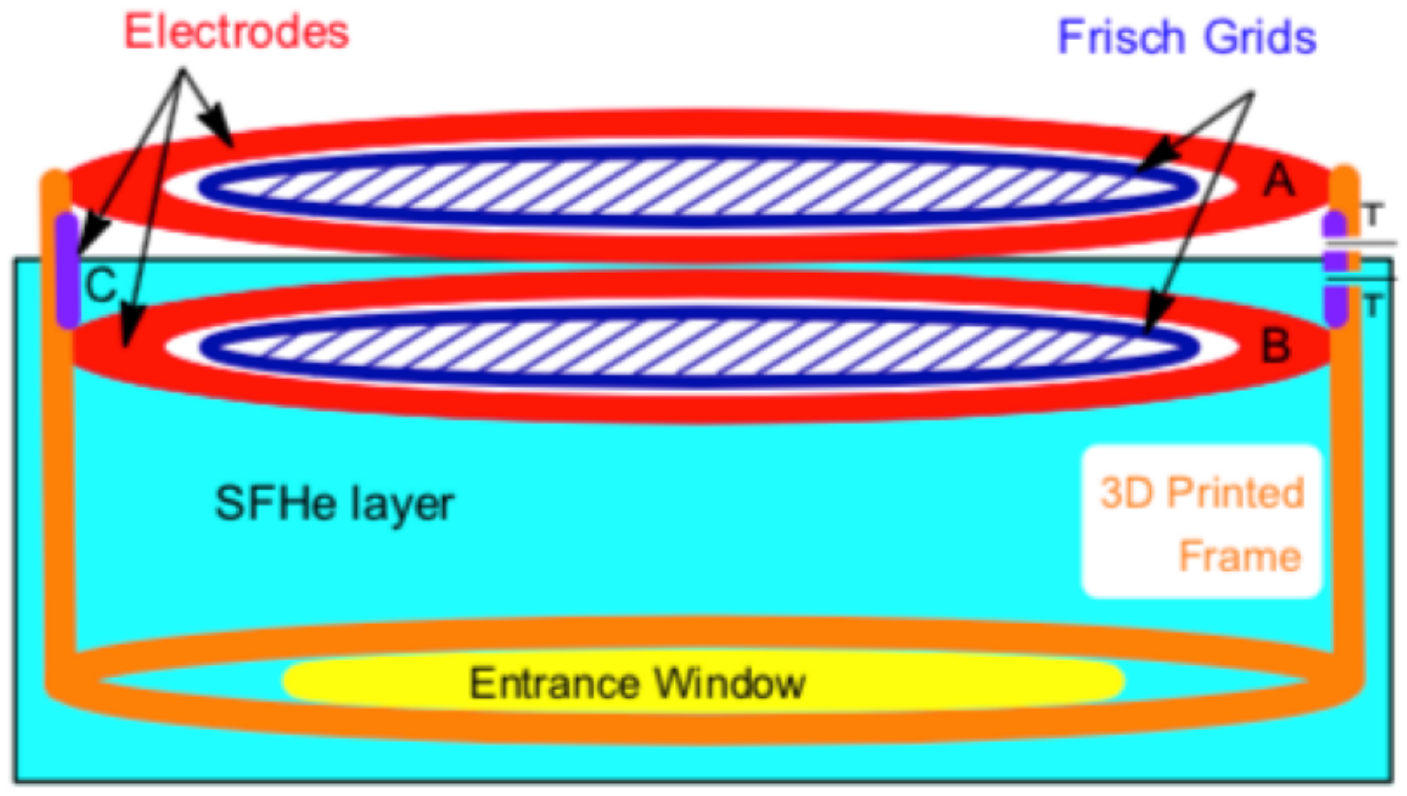}
\caption{\small Conceptual sketch of apparatus to test and characterize charge pool at surface of superfluid helium layer, which creates drift field in helium. (A), (B), (C): field electrodes; (T): tungsten electron-injection tips.}
\label{fig:charge-app}
\end{center}\vspace{-.25in}
\end{wrapfigure}

As a condition to embark on the gravity measurement, we must first demonstrate that Taqqu's proposed muonium-beam formation method works as predicted. We are greatly aided in this by the recent installation of a low-energy $\pi/\mu$ beamline in the Fermilab MTA. Designed to produce 40\,MeV/$c$ muons for a muon-catalyzed fusion ($\mu$CF) experiment, it can easily be tuned for efficient transport of ($\stackrel{<}{_\sim}$\,29\,MeV/$c$) surface muons, as we have shown in simulation. Our first task, which can proceed as soon as the $\mu$CF experiment allows, is to verify these simulations by characterizing the $\mu^+$ spectrum experimentally as a function of magnet currents. Our second task (which can start in parallel with our first) is to demonstrate sufficient control of the SFHe charge-pool technology to maintain the needed drift field in the helium (Fig.~\ref{fig:charge-app}).  This can be done using an existing IIT cryostat, vacuum pumps, and cryocoolers, preferably at Fermilab. The total scale of the project is a few M\$, about half of which is personnel costs.

\vspace{-.15in}\section{Draft Schedule}\vspace{-.05in}

We anticipate that this program will require 3--5 years in total. Phase I: Use G4beamline to simulate the muon beam, optimize its parameters for muonium production in superfluid helium, and characterize the MTA surface-muon beam experimentally;  test and characterize the electric-field-in-helium apparatus;  design the detectors needed to characterize the produced muonium. Phase II: Obtain a suitable dilution refrigerator to reach the $<1$\,K temperature range at which the process works best, build the detectors, install the apparatus within the sample volume, measure muonium production, and determine the optimal operating parameters and maximum muonium yield.  Phase III: Build the additional apparatus needed to observe and measure the effect of the earth's gravity on muonium and carry out a first measurement.


\vspace{-0.15in}\begin{thebibliography}{99}\vspace{-.05in}\small
\setlength{\itemsep}{-0.08em}
\bibitem{Petrov}
A. A. Petrov, R. Conlin, C. Grant, ``Studying $\Delta L=2$ Lepton Flavor Violation with Muons," Universe {\bf 8} (2022) 169, \url{https://www.mdpi.com/2218-1997/8/3/169}. 

\bibitem{Willmann}
L. Willmann {\it et al.}, ``New Bounds from a Search for Muonium to Antimuonium Conversion," Phys.\ Rev.\ Lett.\ {\bf 82} (1998) 49.

\bibitem{Meyer}
V. Meyer {\it et al.}, ``Measurement of the 1s--2s Energy Interval in Muonium,'' Phys. Rev. Lett. {\bf 84} (2000) 1136.

\bibitem{Liu}
W. Liu {\it et al.}, ``High Precision Measurements of the Ground State Hyperfine Structure Interval of Muonium and of the Muon Magnetic Moment," Phys. Rev. Lett. {\bf 82} (1999) 711. 

\bibitem{Antognini}
A. Antognini {\it et al.}, ``Studying Antimatter Gravity with Muonium," Atoms {\bf 6} (2018) 17; 

\bibitem{EotWash-LLR-MICROSCOPE}
T. A. Wagner, S. Schlamminger, J. H. Gundlach, E. G. Adelberger, ``Torsion-balance tests of the weak equivalence principle," Class. Quantum Gravity {\bf 29}  (2012) 184002;
J. G. Williams, S. G. Turyshev, D. H. Boggs, ``Progress in Lunar Laser Ranging Tests of Relativistic Gravity,'' Phys. Rev. Lett., {\bf 93}  (2004) 261101;
P. Touboul {\it et al.} [MICROSCOPE Collaboration], ``MICROSCOPE Mission: Final Results of the Test of the Equivalence Principle," Phys. Rev. Lett. {\bf 129} (2022) 121102.

\bibitem{NG}
M. M. Nieto, T. Goldman, ``The Arguments Against `Antigravity' and the Gravitational Acceleration of
Antimatter," Phys. Rep. {\bf 205} (1991) 221--281.

\bibitem{Chardin}
G. Chardin, ``Motivations for Antigravity in General Relativity," Hyp. Int. {\bf 109} (1997), 83; A. Benoit-L\'{e}vy, G. Chardin, ``Introducing the Dirac--Milne universe," Astron. \& Astrophys. {\bf 537} (2012) A78; G. Chardin {\it et al.}, ``MOND-like behavior in the Dirac--Milne universe: Flat rotation curves and mass versus velocity relations in galaxies and clusters," A\&A {\bf 652} (2021) A91; G. Chardin, ``Experimental and Observational Tests of Antigravity," arXiv:2210.03445 [astro-ph.CO] (2022).

\bibitem{Aaij}
R. Aaij {\it et al.} (LHCb Collaboration), ``Test of lepton universality in beauty-quark
decays," Nat. Phys. {\bf 18} (2022) 277 and references therein.

\bibitem{Abela} 
R. Abela {\it et al.}, ``Muonium in liquid helium isotopes," JETP Lett. {\bf 57} (1993) 157.

\bibitem{Taqqu}
D. Taqqu, ``Ultraslow Muonium for a Muon beam of ultra high quality," Phys. Procedia  {\bf 17} (2011) 216.

\bibitem{Luppov}
V. G. Luppov {\it et al.}, ``Focusing a Beam of Ultracold Spin-Polarized Hydrogen Atoms with a Helium-Film-Coated Quasiparabolic Mirror," Phys. Rev. Lett. {\bf 71} (1993) 2405.

\end{thebibliography}
\end{document}